\documentstyle [12pt]{article}
\begin{document}
  
\hfill {WM-98-118}
  
\hfill {\today}
  
\vskip 1in   \baselineskip 24pt{\Large 
\bigskip
\centerline{Vacuum Stability Bounds in the Two-Higgs Doublet Model}}
   
\vskip .8in
\def\bar{\overline}
\def\lamda{\lambda}
  
\centerline{Shuquan Nie and Marc Sher }
\bigskip
\centerline {\it Nuclear and Particle Theory Group}
\centerline {\it Physics Department}
\centerline {\it College of William and Mary, Williamsburg, VA
23187, USA} 
  
\vskip 1in
  
{\narrower\narrower In the standard model, the requirements of vacuum stability
and the validity of perturbation theory up to the unification scale force the
mass of the Higgs boson to be approximately between 130 GeV and 180 GeV.  We
re-examine these requirements in the (non-supersymmetric) two-Higgs doublet
model, in the light of the large top quark mass,  and constrain the masses of the
Higgs bosons in this model.  It is found that the mass of the charged Higgs
boson must be lighter than 150 GeV.   This bound is below the lower bound in the
popular model-II two-Higgs doublet model, and thus we conclude that this model
cannot be valid up to the unification scale.  The bounds on the neutral Higgs
scalars are also discussed.  }
  
\newpage

The mass of the Higgs boson in the standard model is, at first sight,
completely arbitrary.  It depends on the scalar self-coupling, $\lamda$, which
is a free parameter.  However, rather stringent bounds on the mass can be
obtained\cite{bounda,bounds} by requiring that (a) $\lamda$ remain perturbative
up to a large scale (generally taken to be the unification scale of approximately
$10^{16}$ GeV--the precise value doesn't much matter), and (b) that the vacuum
of the standard model remain stable up to that large scale.  The first condition
gives an upper bound to the Higgs mass of approximately $180$ GeV.  The second
condition is virtually identical to requiring that $\lambda$ remain positive up
to the large scale, and that gives a lower bound to the Higgs mass of
approximately $130$ GeV.  Thus, these two conditions strongly constrain the
mass of the Higgs boson to be between $130$ and $180$ GeV.

It is easy to see where these constraints arise.  The renormalization group
equation for the scalar self-coupling is of the form ${d\lamda\over
 dt}=a\lamda^2-bg^4_Y$, where $g_Y$ is the top quark Yukawa coupling.  If
$\lambda$ is large, the first term dominates, and $\lambda$ blows up; if it is
small, the second term dominates, and $\lambda$ becomes negative, leading to a
vacuum instability.  Only for $\lambda$ near the fixed point of this equation
does $\lambda$ remain positive and finite from the electroweak to unification
scale.
  
  The most straightforward extension
  of the standard model is the two-Higgs doublet model(2HDM). It includes
two complex scalar doublets (see Ref. \cite{bounda} for a review),
    
  \begin{equation}
  \Phi_1=
  \left (
  \begin{array}{c}
  \chi_{1}^{+} \\
  (\phi_1+i\chi_1)/\sqrt{2}
  \end{array}
  \right )
  ,\qquad
  \Phi_2=
  \left (
  \begin{array}{c}
  \chi_{2}^{+} \\
  (\phi_2+i\chi_2)/\sqrt{2}
  \end{array}
  \right )
  \end{equation}
  Of the eight real fields, three must become the longitudinal components
  of the $W^{  \pm}$ and Z bosons after the spontaneous symmetry breaking.
  Linear combinations of $\chi_1^{  \pm}$ and $\chi_2^{  \pm}$ are absorbed
  into the longitudinal parts of the $W^{  \pm}$ bosons and a linear 
  combination of $\chi_1$ and $\chi_2$ give mass to the Z boson. Five physical
  Higgs scalars will remain: a charged scalar $\chi^{  \pm}$ and three neutral 
  scalars $\phi_1$, $\phi_2$ and the other linear combination of $\chi_1$
  and $\chi_2$, called $\chi^0$.

	Given how stringently the Higgs mass in the standard model is constrained, one
might ask how stringently the masses of the scalars in the two-doublet model
are constrained.  There are many more self-couplings (which could potentially
diverge by the unification scale) and more directions in field space where an
instability could arise.  In this paper, we examine these constraints in the
two-doublet model.   This is not new---the constraints have been examined
before\cite{bounda,boundb}, but the top quark mass was unknown at the time (and
only values up to about $130$ GeV for the top quark mass were considered).
  
  A potential danger with additional Higgs doublets  is the possibility of 
  flavor-changing neutral currents(FCNC). It is well known that FCNC are highly 
  suppressed relative to the charged current processes, so it would be desirable
  to ``naturally" suppress them in these models. If all quarks with the same
quantum 
  numbers couple to the same scalar multiplet, then FCNC will be absent. This led
  Glashow and Weinberg\cite{glashowweinberg} to propose a discrete symmetry which
force all the quarks
  of a given charge to couple to only one doublet. There are two such possible 
  discrete symmetries in the 2HDM,
  \begin{equation}
  (I) \phi_2\rightarrow -\phi_2\qquad\qquad (II)\phi_2
 \rightarrow -\phi_2, d_R^i \rightarrow -d_R^i 
\end{equation}   
In model I, all quarks couple to the same doublet, and no
quarks couple to the other
  doublet. In model II, the Q=2/3 quarks couple to one doublet and the Q=-1/3
quarks couple to the other
  doublet.

  The most general potential subject to one of the discrete symmetries, for two
doublets of hypercharge 
  +1 (if one of the doublets has hypercharge -1, our arguments will be
unaffected), is
   \begin{eqnarray}
 V&=&\mu_1^2\Phi_1^{+}\Phi_1+\mu_2^2\Phi_2^{+}\Phi_2+
\lamda_1(\Phi_1^{+}\Phi_1)^2
+\lamda_2(\Phi_2^{+}\Phi_2)^2\cr &+&\lamda_3(\Phi_1^{+}\Phi_1)(\Phi_2^{+}\Phi_2)
 +\lamda_4(\Phi_1^{+}\Phi_2)^2+{1\over
2}\lamda_5[(\Phi_1^{+}\Phi_2)^2+(\Phi_2^{+}\Phi_1)^2]
  \end{eqnarray}  
  the vacuum expectation values of $\Phi_1$ and $\Phi_2$ can be written as
   \begin{equation}
  \Phi_1=1/\sqrt{2}
  \left(
  \begin{array}{c}
  0 \\
  v_1
  \end{array}\right)
  ,\qquad
  \Phi_2=1/\sqrt{2}
  \left(
  \begin{array}{c}
  0 \\
  v_2
  \end{array}\right)
  \end{equation}
  with $v_1^2+v_2^2=\sqrt{2}G_F=(247\ GeV)^2$.
  The masses of the physical scalars are given by
  \begin{equation}
  m_{\chi^{  \pm}}^2=-{1 \over 2}(\lamda_4+\lamda_5)(v_1^2+v_2^2),\qquad
  \qquad
  m_{\chi^{0}}^2=-\lamda_5(v_1^2+v_2^2),
   \end{equation}
\begin{equation}
  m_{\phi}^2={1 \over 2}(A+B+\sqrt{(A-B)^2+4C^2}),\qquad
  \qquad
  m_{\eta}^2={1 \over 2}(A+B-\sqrt{(A-B)^2+4C^2}), 
  \end{equation}
  where $A=2\lamda_1
  v_1^2$, $B=2\lamda_2v_2^2$ and $C=(\lamda_3+\lamda_4+\lamda_5)v_1v_2$.
  
  It is required that all scalar boson masses-squared must be positive.This
implies
  that
  \begin{equation}
  \begin{array}{l}
  (i)\ \lamda_5<0
  \\
  (ii)\ \lamda_4+\lamda_5<0
  \\
  (iii)\ \lamda_1>0
  \\
  (iv)\ \lamda_2>0
  \\
  (v)\ 2\sqrt{\lamda_1\lamda_2}>\lamda_3+\lamda_4+\lamda_5
  \end{array}
  \end{equation}
  
In the standard model the positivity of the scalar boson mass-squared implies
that 
  $\lamda>0$, To ensure vacuum stability for all scales up to $M_X$, one must 
  have $\lamda(q^2)>0$ for all $q^2$ from $M_Z^2$ to $M_X^2$.
  Similarly, to ensure vacuum stability in the 2HDM up to $M_X^2$, one must 
  require that all of the five constraints be valid up to $M_X^2$. If the 
  condition $\lamda_1$ or $\lamda_2>0$ is violated, the potential will be
unstable
  in the $\phi_1$ or $\phi_2$ direction. if the condition ($v$) is violated, the 
  potential will be unstable in some dirction in $\phi_1-\phi_2$ plane. If
  the condition $\lamda_4+\lamda_5<0$ is violated, a new minimum which breaks
  charge will be formed. And if $\lamda_5$ become positive, a new minimum
  which violates CP will be formed. Thus it is required that all of the 
  constraints should be valid up to $M_X$. At the same time it is physically 
  reasonable to demand that all $\lamda$'s be finite (or perturbative) up
to $M_X$.     
  
  Starting with $\lamda_1,\lamda_2,\lamda_3,\lamda_4,\lamda_5$ and $\tan\beta$
at the electroweak scale,
  the renormalization group equations are integrated numerically to check whether
  one of the five constraints is violated or whether any of the couplings become
nonperturbative before reaching  
  $M_X$.   The results give an allowed region in the six-dimensional parameter
space.   To explore this region, we choose the six parameters to be the four
physical scalar masses, $\tan\beta$ and $\lamda_3$.  For a point in this
parameter-space to be acceptable, all of the above constraints, as well as
perturbativity, must be satisfied at all scales up to $M_X$.  In practice, since
$\lambda_3$ is unmeasureable, we consider the other five parameters as starting
points, and see if any initial values of $\lambda_3$ give acceptable values. 
In this way, we determine if a given point in the five-dimensional space of the
scalar masses and $\tan\beta$ is acceptable.

It is, of course, difficult to plot a region in five-dimensional space. 
However, the basic features can easily be seen with a few examples.   Let us
first consider the case in which $m_{\chi^\pm}=m_{\chi^o}=100$ GeV.  We choose
$\tan\beta=2$, and plot the allowed region in the neutral scalar mass plane.
The region is shown in Figure 1.   For $m_\eta$ between $40$ and $88$ GeV, one
sees that the value of $m_\phi$ must lie below $180$ GeV and above a value
which ranges from $130$ GeV to $100$ GeV; this bound  is very similar to the
result in the standard model.  However, there are no solutions in which
$m_\eta$ is greater than $88$ GeV or below $40$ GeV.  Below the region,
$\lambda_1$ becomes negative, above the region $\lambda_1$ becomes
non-perturbative, to the left and to the right of the region, the constraint
($v$) is violated.

One can now vary some of the three input parameters to see how this region
changes.   As the charged Higgs mass increases, the region shrinks
dramatically, disappearing when it reaches $140$ GeV, as shown in Figure 2.    As
$\tan\beta$ increases, the region shifts to smaller values of $m_\eta$, as shown
in Figure 3.  This is not surprising since
$m_\eta$ becomes small as $\tan\beta\rightarrow\infty$.  As $\tan\beta$
decreases, the size of the allowed region shrinks, since the top quark Yukawa
coupling is getting larger, leading to an instability.   Finally, varying
$m_{\chi^o}$ gives the result in Figure 4.  As in the charged Higgs case, the
allowed region disappears when the pseudoscalar mass exceeds $140$ GeV.

The most important result is seen from Figure 2, where this is a stringent
upper bound on the charged Higgs mass.  By optimizing $\tan\beta$ and
$m_{\chi^o}$, we find that the maximum allowed value for the charged Higgs mass
is $150$ GeV .  

What are the experimental constraints?  As discussed in Ref. \cite{who}, there
are very few constraints on the neutral scalar masses.  If one takes the $\chi^o$
mass to be $100$ GeV, the only constraints come from the Bjorken process,
$e^+e^-\rightarrow Z^*\rightarrow Z\eta$, and the rate can be significantly
reduced by judicious choice of the mixing angle.  So no bounds on $m_\eta$ are
relevant.  There is, however, a strong bound\cite{who,aliev} on $m_\chi^\pm$
coming from $B\rightarrow X_s\gamma$.  In Model II, this process forces the
charged Higgs mass to be greater than $165$ GeV.    This in inconsistent with
our upper bound.  Model I, however, has no such constraint, and the charged
Higgs mass could be as light as $45$ GeV.

We conclude that the popular two-Higgs doublet model, Model II, can not be
valid up to the unification scale.  Model I is not excluded, however
we do find that the charged Higgs mass must be lighter than $150$ GeV, the
lightest neutral scalar must be lighter than $110$ GeV and the pseudoscalar must
be lighter than $140$ GeV for the model to be valid up to the unification scale.

\def\prd#1#2#3{{\rm Phys. ~Rev. ~}{D#1} (19#2) #3 }
\def\plb#1#2#3{{\rm Phys. ~Lett. ~}{B#1} (19#2) #3  }
\def\npb#1#2#3{{\rm Nucl. ~Phys. ~} {B#1} (19#2) #3}
\def\prl#1#2#3{{\rm Phys. ~Rev. ~Lett. ~}{#1} (19#2)  #3 }

\newpage

\begin{figure}
\caption{The allowed region in the neutral scalar mass plane, with
$m_{\chi^\pm}=m_{\chi^o}=100$ GeV and $\tan\beta=2$.}
\end{figure}

\begin{figure}
\caption{The allowed region in the neutral scalar mass plane for various values
of the charged Higgs mass (in GeV), with the $\chi^o$ mass chosen to be $100$
GeV and $\tan\beta=2$.}
\end{figure}

\begin{figure}
\caption{The allowed region in the neutral scalar mass plane for various values
of $\tan\beta$, with the $\chi^o$ and $\chi^\pm$ masses chosen to be $100$ GeV.}
\end{figure}

\begin{figure}
\caption{The allowed region in the neutral scalar mass plane for various values
of the pseudoscalar Higgs mass (in GeV), with the $\chi^\pm$ mass chosen to be
$100$ GeV and $\tan\beta=2$.}
\end{figure}

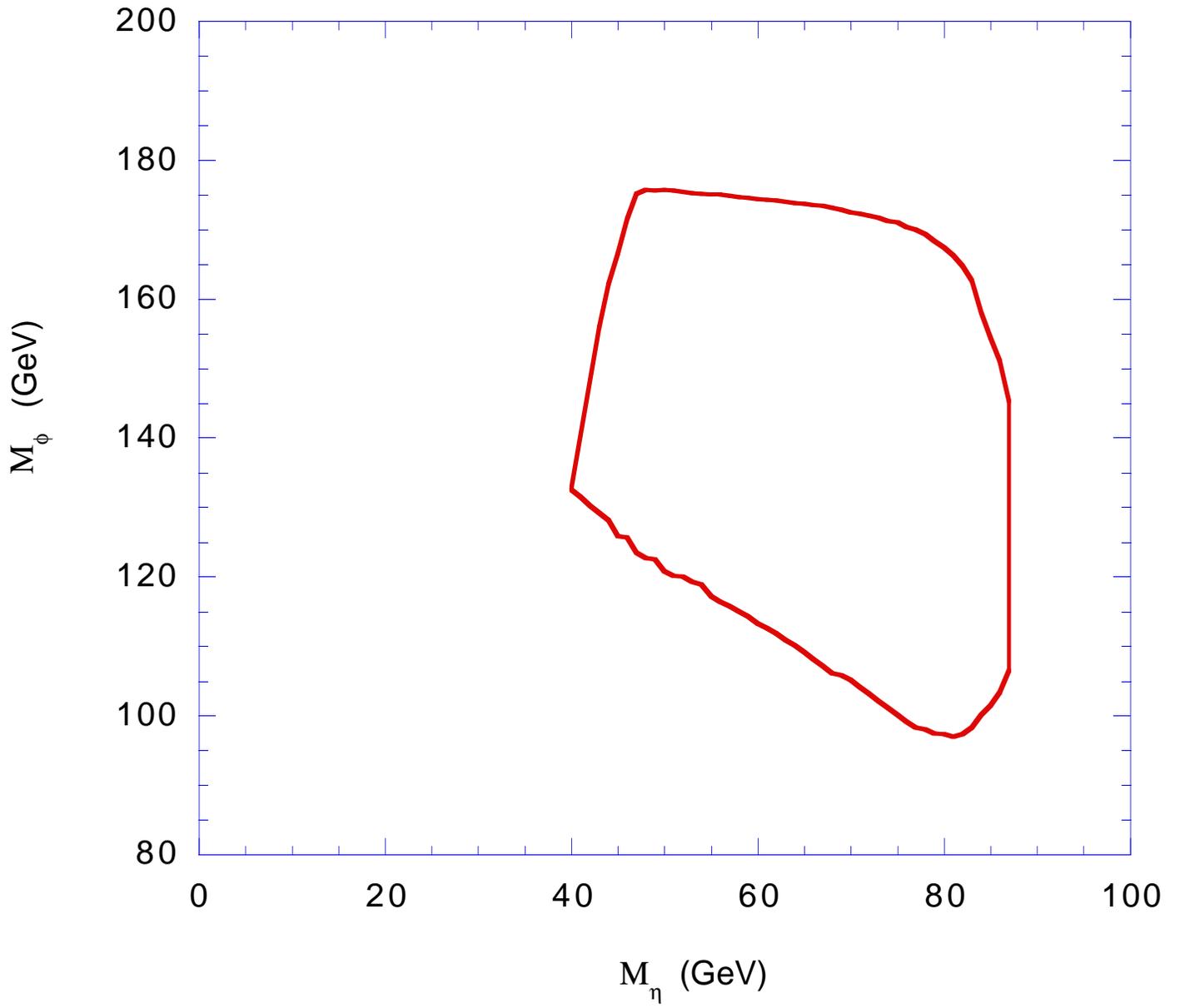
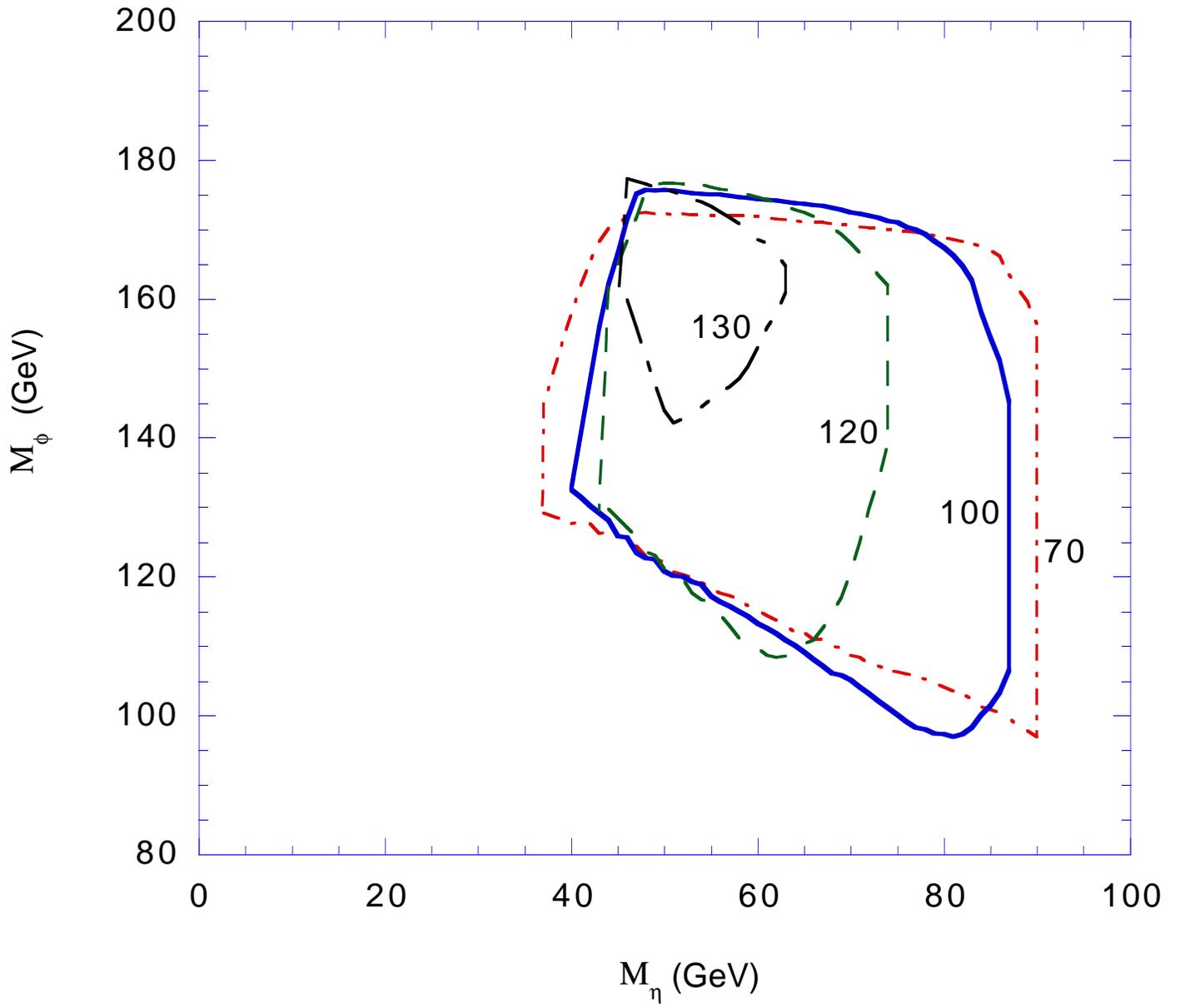
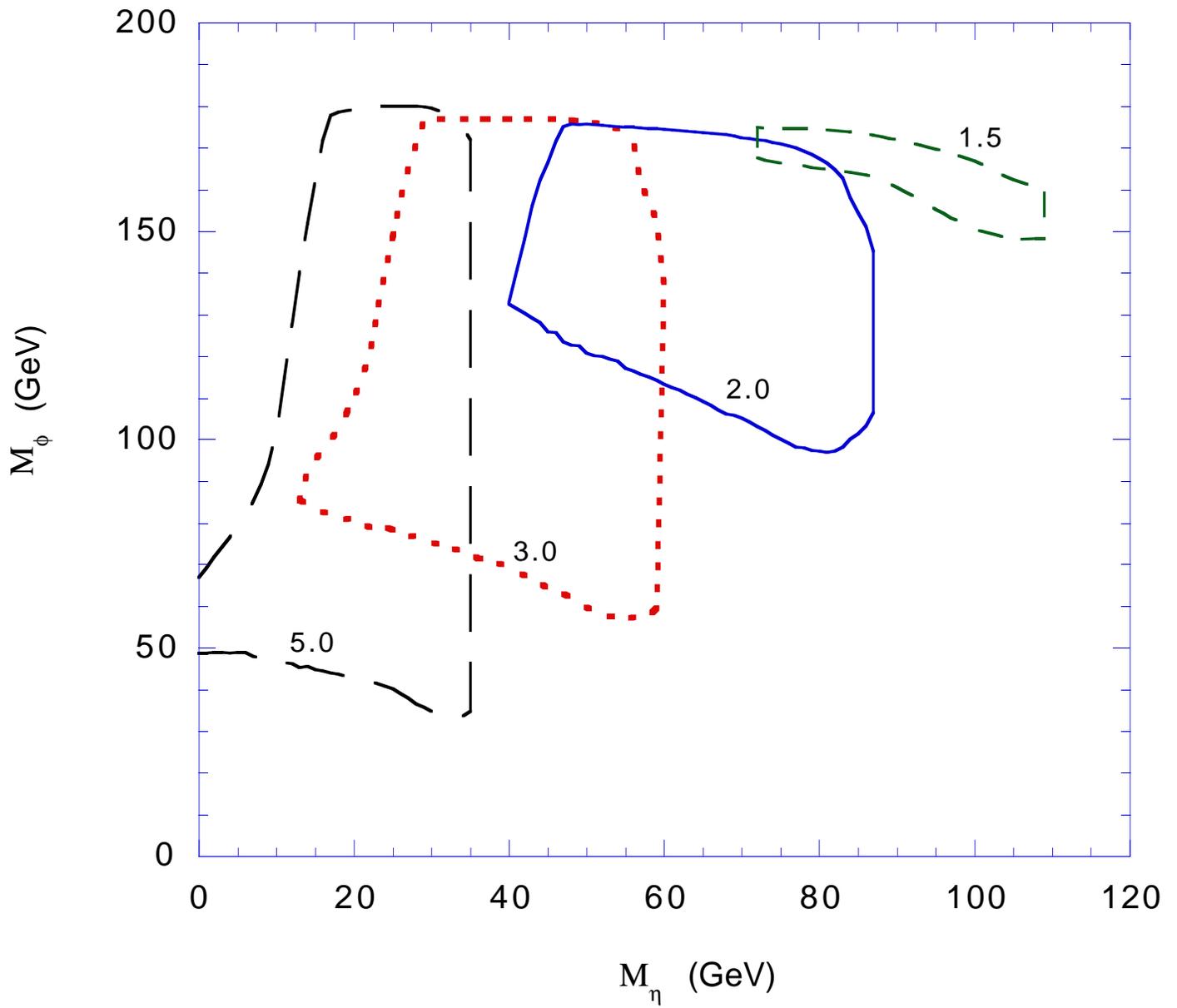
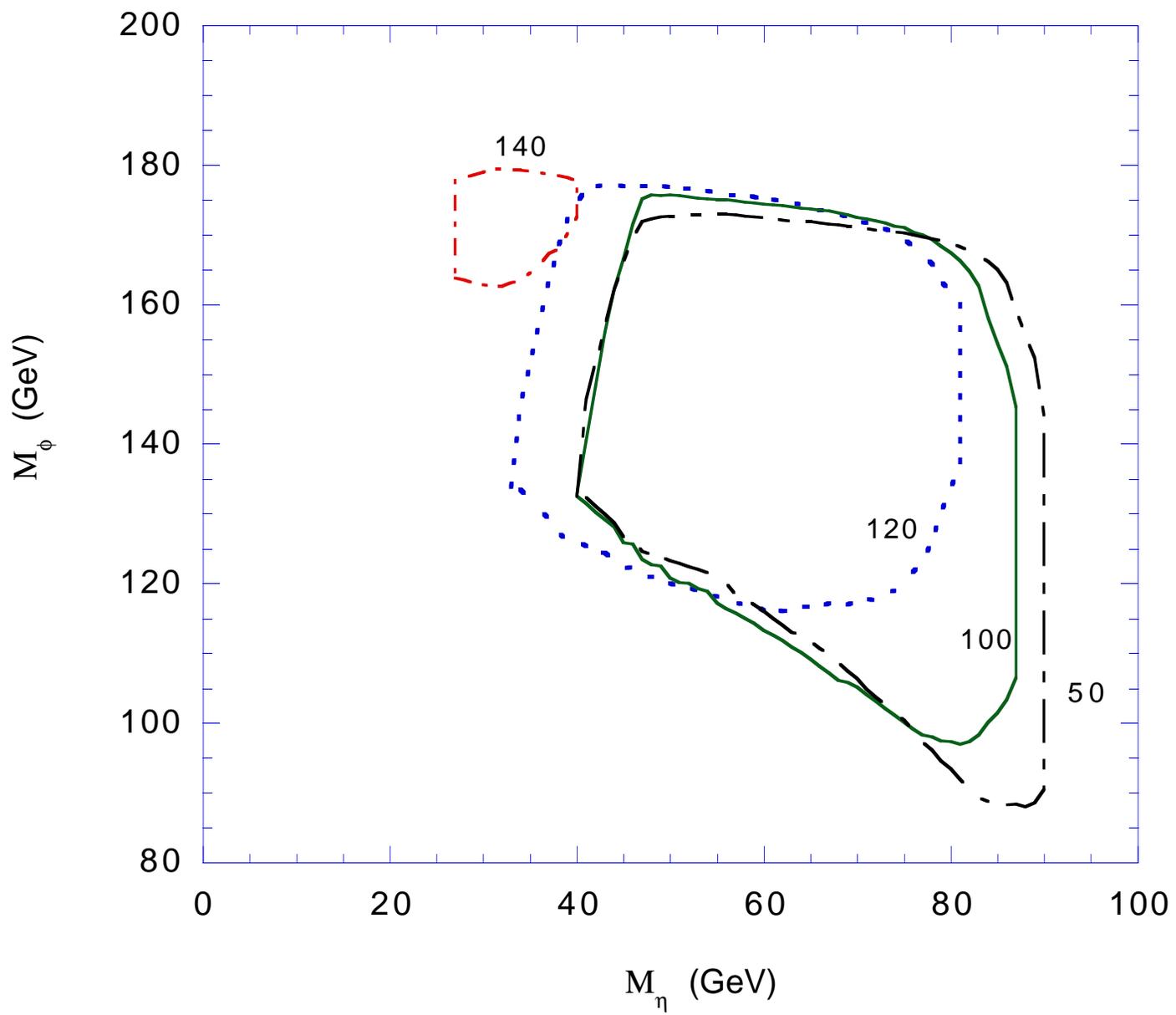
\end{document}